\documentclass[usenatbib]{mn2e}

\newif\ifAMStwofonts \AMStwofontstrue

\ifCUPmtlplainloaded \else
  \ifAMStwofonts \else 
    \def\upi{\pi}
    \def\umu{\mu}
    \def\upartial{\partial}
  \fi
\fi

\title{The symbiotic star CH Cygni. III. A precessing radio jet}

\author[M. M. Crocker et al.]  {M. M. Crocker$^1$,  R. J. Davis$^1$, R. E. Spencer$^1$,
       S. P. S. Eyres$^2$, M. F. Bode$^3$, A. Skopal$^{3,4}$\\
       $^1$Jodrell Bank Observatory, University of Manchester, Macclesfield, Cheshire, SK11 9DL, UK.\\
       $^2$Centre for Astrophysics, University of Central Lancashire, Preston, PR1 2HE, UK\\
       $^3$Astrophysics Research Institute, Liverpool John Moores University, Twelve Quays House, Egerton Wharf, Birkenhead, CH41 1LD, UK\\ 
       $^4$Astronomical Institute, Slovak Academy of Sciences, 059~60 Tatransk\'{a} Lomnica, Slovakia
}

\date{Accepted ????.  Received ????; in original form ????}

\pagerange{\pageref{firstpage}--\pageref{lastpage}} \pubyear{2001}

\begin{document}

\maketitle

\label{firstpage}

\begin{abstract}

VLA, MERLIN and Hubble Space Telescope imaging observations of the
extended regions of the symbiotic system CH Cygni are analysed.
These extensions are evidence of a strong collimation mechanism,
probably  an accretion disk surrounding the hot component of the
system. Over 16 years (between 1985 and 2001) the general trend is
that these jets are  seen to precess. Fitting a simple ballistic model
of matter ejection to the geometry of the extended regions suggests a
period of  $6520 \pm 150$ days, with a precession cone opening angle
of $35 \pm 1$ degrees. This period is of the same order as that
proposed for the orbital period of the outer giant in the system,
suggesting a possible link between  the two. Anomalous knots in the
emission, not explained by the simple model, are believed to be the
result of older, slower moving ejecta, or possibly jet material that
has become disrupted through sideways interaction with the
surrounding medium.

\end{abstract}


\section{Introduction}

Symbiotic stars occupy an extreme and relatively poorly understood
region of the binary classification scheme. The name was coined by
\citet{PWM41} to describe stars which appeared to  have a combination
spectrum: that of high excitation lines usually associated with a hot,
ionised nebula superimposed on a cool continuum with prominent
absorption features consistent with a late--type star. At present they
are understood to be interacting binaries (with orbital periods of a
few to tens of years) consisting of a cool giant losing material
mostly via the stellar wind and a hot, luminous compact object which
ionises a portion of the cool component wind \citep{ERS84}.  Such a
state  of affairs represents the so--called {\em quiescent phase},
which can be interrupted by periods of activity. The {\em active
phases} start with an eruption of the hot star, an event indicated
photometrically by an increase of the star's brightness by 2-6\,mag,
and/or spectroscopically by high velocity broad emission features from
the central star. Both radio and Hubble Space Telescope (HST) imaging 
can directly resolve the remnants of such dramatic events (\citealt{WJH93}; 
\citealt{SPSE95}; \citealt{HTK96}; \citealt{AMSR99}; \citealt{SPSE01a}).

The symbiotic star CH\,Cygni displays particularly complex
behaviour. Optical spectroscopic studies by \citet{MM87} showed that
the orbits of the stars within the system are likely to be coplanar
and eclipsing, with eclipses separated by around 5700 days.  Later
studies (\citealt{MM90b} and references therein) confirmed this period.

Further spectral studies of the system \citep{KHH93} suggested that
CH\,Cyg is probably a triple-star system consisting of an inner
756-day period binary which is orbited by an unseen G-K dwarf on a
5300~day orbit. \citet{AS96b} uncovered 756 day interval eclipses
which show that all three stars in the system are likely to be in
coplanar orbits.

Each of the three outbursts seen since 1978 was accompanied by 
high velocity broad emission features
consistent with mass outflows. During 1984-85 the material was ejected
at $\sim$600-2500\,km\,s$^{-1}$ \citep{MM86}, whilst the 1992--95
active phase was characterized  by sporadic and in part bipolar
outflow at $\sim$1000-1600\,km\,s$^{-1}$ (\citealt{LL96}; \citealt{AS96a};
\citealt{TI96}) and finally, during the recent, 1998--2000, outburst
mass outflows at about 1000\,km\,s$^{-1}$ were observed
(\citealt{SPSE01b},  hereafter Paper II). The outflows may be
correlated with a significant increase of the radio emission and the
radio light curves during these periods fit well with the
optical ones \citep{HTK96}. The 1984 mass ejection has been linked to
the emergence of bipolar emission \citep{ART86} which has been attributed
to high velocity ($\sim$1000\,km\,s$^{-1}$) jets. Non--thermal
emission features have  been discovered in these jets from radio
observations (\citealt{MMC01},  hereafter Paper I), explained by
shocked regions that arise when the high--speed ejecta interact with
existing wind material.

This paper aims to extend our understanding of the jets in this
interesting system by developing theories and models to explain the
change in their morphology in the years following their emergence.

In this paper we present analysis of radio data from the VLA and MERLIN, along
with HST observations.

\section{Observations}

CH Cygni has been observed on several occasions since 1985, in several
different radio wavelengths and optical domains, further details of
which are given in Paper I. All observations have been presented previously
with the exception of the 2000 November 10? VLA data which is shown here
for the first time.

Radio data reduction was carried out using the AIPS software package
\citep{EG99}. The flux density of the phase reference source was
determined, and the data calibrated to correct phase and amplitude
instabilities. These corrections were applied to the observations of
CH Cygni and a map produced by deconvolving the instrumental response
(the so-called ``dirty beam'') from the map using the well-established
CLEAN algorithm \citep{JAH74}.

Observations made on 1999 August 12 with the Hubble Space Telescope's
Wide Field and Planetary Camera 2 through the F502N filter, corresponding 
to the [O III] line at $\lambda\lambda$4956 and 5007\r{A}) \citep{JB01},
are used here as these, with a diffraction limted resolution of around 
0.05~arcseconds, offer direct comparison with the spatial resolution 
of the radio data (see Table \ref{tab:restoringbeam}.  Each observation 
was made up of two sub--exposures, allowing cosmic ray subtraction to be 
carried out (see Paper II and references therein).

\section{Initial data analysis}
\label{sec:precinner}

A preliminary examination of radio maps of CH Cygni (taken with the
VLA and MERLIN), as well as the HST optical images (using the F502N
[OIII] filter) show that the position angle of the extended regions
changes drastically over the 15 years between the earliest (1985) and
most recent (2000) observations.

A more quantitative analysis requires a measurement of the position
angle of the central regions of the radio nebula. A problem inherent
in this method is that the geometry of the restoring beam (in the case
of the  radio observations) may dominate over any natural geometry of
the region.  This arises as a result of the $(u,v)$ coverage of the
instrument, which depends upon the altitude of the source above the
horizon.

To avoid these problems, the radio maps were recreated using the
methods of \citet{JAH74} but a circular restoring beam was imposed
upon them, as laid out in Table \ref{tab:restoringbeam}.

\begin{table} 
\caption{Restoring beams used at different radio observing
frequencies.}
\begin{tabular}{|l |l |l |} 
\hline  
Instrument  &Frequency (GHz) &Beamsize (arcseconds)\\  
\hline
VLA         &1.5             & 1.5\\
            &4.8             & 0.5\\
            &8.4             &0.25\\  
            &15.0            & 0.15\\
MERLIN      &1.5             &0.15\\  
            &4.9             & 0.05\\  
\hline
\end{tabular}  
\label{tab:restoringbeam} 
\end{table}

Two dimensional Gaussians were fitted to the bright peak in the centre
of each map, and the position angle (anticlockwise through east from
north) of the semi--major axis of the resulting ellipse was taken as
the position angle of the central material. The uncertainties are those
produced by the AIPS task 'IMFIT', and are derived from the ratio of the
axes of the fitted ellipse. These measurements are shown in Table \ref{tab:pa}.

\begin{table} 
\caption{Position angle of the central region of the nebula of CH~Cygni}
\begin{tabular}{|l |l |l |l |l|} 
\hline  Instrument &Julian Date   &Year &$\nu$ (Ghz)&Position angle ($^\circ$)\\
\hline  
VLA                &2446087 &1985 &14     &$137.7 \pm 0.3$\\  
VLA                &2446509 &1986 &1.6    &$135.4 \pm 1.2$\\  
VLA                &2446509 &     &4.8    &$133.9 \pm 0.3$\\  
VLA                &2446509 &     &14     &$123.2 \pm 1.6$\\  
VLA                &2446509 &     &22     &$131.1 \pm 10.7$\\  
VLA                &2446576 &     &1.6    &$136.7 \pm 1.0$\\  
VLA                &2446576 &     &4.8    &$132.2 \pm 0.4$\\  
VLA                &2446576 &     &14     &$117.2 \pm 0.7$\\    
VLA                &2446576 &     &22     &$113.3 \pm 4.4$\\    
VLA                &2447003 &1987 &1.5    &$134.9 \pm 2.0$\\  
VLA                &2447003 &     &4.8    &$110.5 \pm 0.5$\\    
VLA                &2447003 &     &15     &$91.8 \pm 1.2$\\  
VLA                &2447450 &1988 &1.6    &$138.9 \pm 5.6$\\  
VLA                &2447450 &     &4.8    &$109.8 \pm 4.5$\\
VLA                &2447505 &     &14     &$111.7 \pm 20.6$\\
VLA                &2448488 &1991 &4.8    &$112.0 \pm 9.7$\\    
MERLIN             &2448827 &1992 &4.8    &$121.0 \pm 9.9$\\  
MERLIN             &2449911 &1995 &4.8    &$125.5 \pm 5.0$\\ 
VLA                &2449916 &     &1.6    &$138.9 \pm 4.8$\\  
VLA                &2449916 &     &8.4    &$132.0 \pm 2.0$\\  
VLA                &2449916 &     &14     &$138.0 \pm 16.8$\\  
VLA                &2449949 &     &1.6    &$139.5 \pm 3.1$\\  
VLA                &2449949 &     &4.8    &$134.9 \pm 1.4$\\  
VLA                &2449949 &     &8.4    &$126.7 \pm 1.1$\\  
MERLIN             &2451235 &1999 &4.8    &$174.3 \pm 4.7$\\    
MERLIN             &2451307 &     &1.7    &$166.5 \pm 7.2$\\   
HST                &2451402 &     &[OIII] &$162.2 \pm 0.1$\\  
VLA                &2451447 &     &4.8    &$164.1 \pm 0.7$\\  
VLA                &2451447 &     &8.4    &$169.9 \pm 0.8$\\  
MERLIN             &2451481 &     &1.6    &$166.5 \pm 7.2$\\  
MERLIN             &2451607 &2000 &4.8    &$200.5 \pm 5.2$\\  
MERLIN             &2451696 &     &4.8    &$185.4 \pm 8.0$\\   
VLA                &2451858 &     &1.6    &$177.2 \pm 5.2$\\  
VLA                &2451858 &     &4.8    &$171.5 \pm 1.4$\\  
VLA                &2451858 &     &8.4    &$192.2 \pm 1.0$\\  
VLA                &2451858 &     &14     &$197.6 \pm 10.3$\\  
MERLIN             &2451961 &     &4.8    &$198.6 \pm 18.7$\\   
\hline
\end{tabular}  
\label{tab:pa} 
\end{table}

\begin{figure*}
\begin{picture}(600,300)
\put(0,0) {\includegraphics{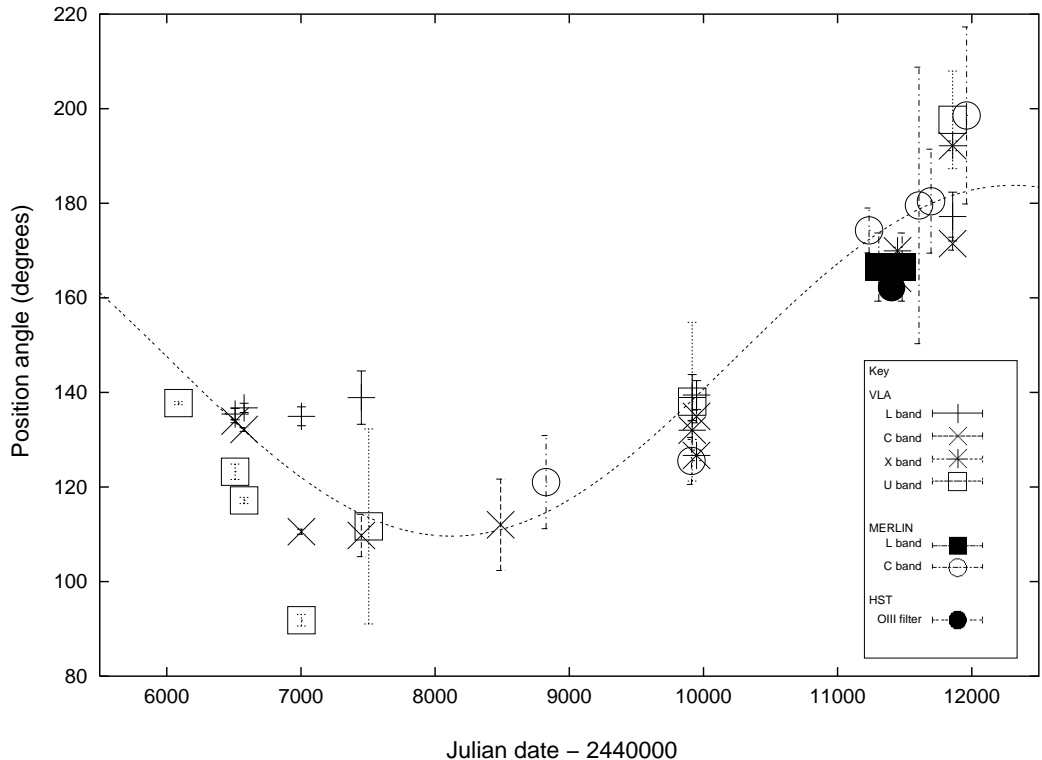}}
\end{picture} 
\caption{Variation of the position angle of the central region of CH
Cygni between 1985 and 2000, along with the fitted curve described
in Section \ref{sec:precinner}. The fitted curve should be treated
with caution as the data set does not appear to cover a full period
of precession.}
\label{fig:pacurve}
\end{figure*}

Since the details of the alignment of the CH Cygni system are not
known, it was assumed that the orbital plane of the system is being
viewed edge--on, as eclipses are seen within the system
(\citealt{MM87}; \citealt{AS96b}). The extension might be expected to 
precess about an axis perpendicular to the orbital plane if this is 
associated with the collimating mechanism. Assuming that the rate 
of precession is constant, the position angle, $\theta_{\rm PA}$ should  
follow a simple sinusoid so that at time $t$ the position angle is 
given by

\begin{equation}
\theta_{\rm PA} = \theta_{\rm mid}+\theta_{\rm o}{\rm
cos}\Biggl(2\pi\times\frac{t-t_0}{P}\Biggr).
\end{equation}

\noindent Such a curve was fitted, giving a precession of  period $P =
8373 \pm 800$ days about a position angle on the sky of  $\theta_{\rm
mid} = 146.7 \pm 3.4$ degrees. The opening angle of the precession
cone was $\theta_{\rm o} = 37.1 \pm 4.7 ^\circ$. 

This precession period is longer than the $5294 \pm 117$ day period
quoted for the length of the orbit of the outer giant of the system by
\citet{KHH93}. The precession period we see could be a resonance of
this, the 756 day orbital period, or both. However, the period we find
should be treated with some caution, as our data set does not appear to
cover a full period of precession.

\section{Evidence of precession within the radio jets}

There is a large deviation of some of the points from the fitted
curve. A flaw inherent in the method of fitting gaussians is that
it will naturally pick up the position angle of material at a certain
distance from the core -- that distance being determined by the
resolution of the map. Hence what is being picked up is the position
angle of material that was ejected some time before the observation
was made.

If the material in the extension is being ejected ballistically from a
precessing source then it would be expected to trace out a conical
helix of a form seen in many precessing objects \citep{LZ88}. Expansion
has been  detected in the jet since the first radio jets were seen by
\citet{ART86}.

Precessing jets are frequently seen in astronomical objects, one of
the best studied being SS433. Its extended nebula exhibits a
characteristic ``S'' profile that has been attributed to a precessing
jet \citep{RMH81} with great success. Code derived from the work of 
R. E. Spencer (priv. comm.) was adapted for CH Cygni by removing the 
relativistic behaviour of the jets. This code gives the on--the--sky 
positions of discrete knots in the jet at a given time, as determined
by a set of parameters. These patterns were compared to the distribution
of the 40 brightest CLEAN components in each map at a given frequency 
and the deviation minimised through use of the Downhill Simplex Method 
\citep{JAN65}. 

The process was carried out using the set of 5 GHz maps taken by the 
VLA in the highest resolution ``A'' configuration, as these provided 
the most extensive coverage of the system over time. 

\subsection{Simple ballistic model}
\label{sec:ballistic}

\begin{figure}
\begin{picture}(200,300)
\put(0,0) 
{\includegraphics{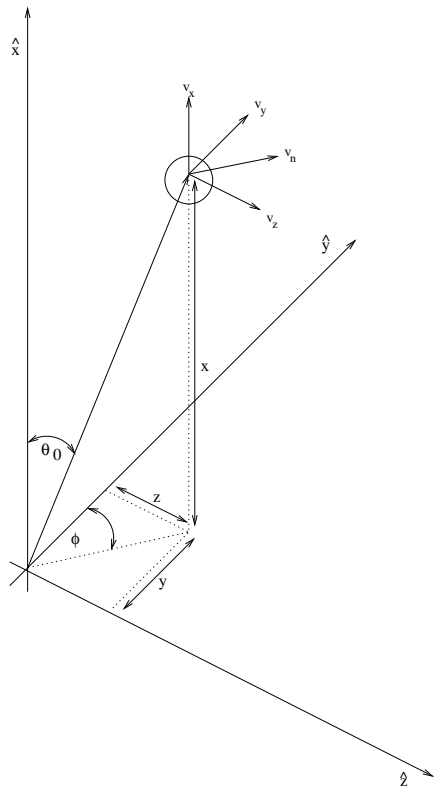}} 
\end{picture} 
\caption{Coordinate system used to describe a simple precessing jet.
The hot component is at the origin, and the jet is represented by
discrete packets of material with coordinates $(x,y,z)$.}
\label{fig:simplejet}
\end{figure}

The simplest model of radio jet precession is one in which material is
ejected continuously, at a constant velocity, from the poles of a
precessing object. The locus described by the points along the
resulting curved path at a time $t$ is described (see Figure 
\ref{fig:simplejet}) in parametric form by

\begin{equation}
x = v_0(t-\tau)~{\rm cos}\theta_0
\label{eq:xjet}
\end{equation}
\begin{equation}
y = v_0(t-\tau)~{\rm sin}\theta_0~{\rm sin}(\phi)
\label{eq:yjet}
\end{equation}
\begin{equation}
z = v_0(t-\tau)~{\rm sin}\theta_0~{\rm cos}(\phi)
\label{eq:zjet}
\end{equation}

\noindent where $x$, $y$ and $z$ are the coordinates of a given knot
of material, ejected from the central source (at the origin) at a
time $\tau$ with velocity $v_0$. The opening angle of the precession 
cone is $\theta_0$, measured from the $x$ axis, and the azimuthal angle 
of the jet (relative to the $y$ axis) is $\Phi = \Omega\tau$ where $\Omega$ 
is the precession rate.

Such a model has been used to fit curves to the high--velocity jets of
the microquasar SS433 by \citet{RES84}. For this model, the parameters 
were: $P$, the precession period of the
jet ($=\frac{2\pi}{\Omega}$); $\tau_0$, a fixed time corresponding to a
point when the  precession angle of the jets was along the line of
sight (the $y$ axis in Figure \ref{fig:simplejet}); $\theta_{\rm mid}$
the position angle (on the sky) of the axis of precession; $\theta_0$,
the opening angle of precession cone; $v_0$, the velocity of the material
ejected in the jet and $i$, the inclination angle of the precession
axis to the line of sight.

\begin{figure*}
\begin{picture}(400,600)
\put(0,0) 
{\includegraphics{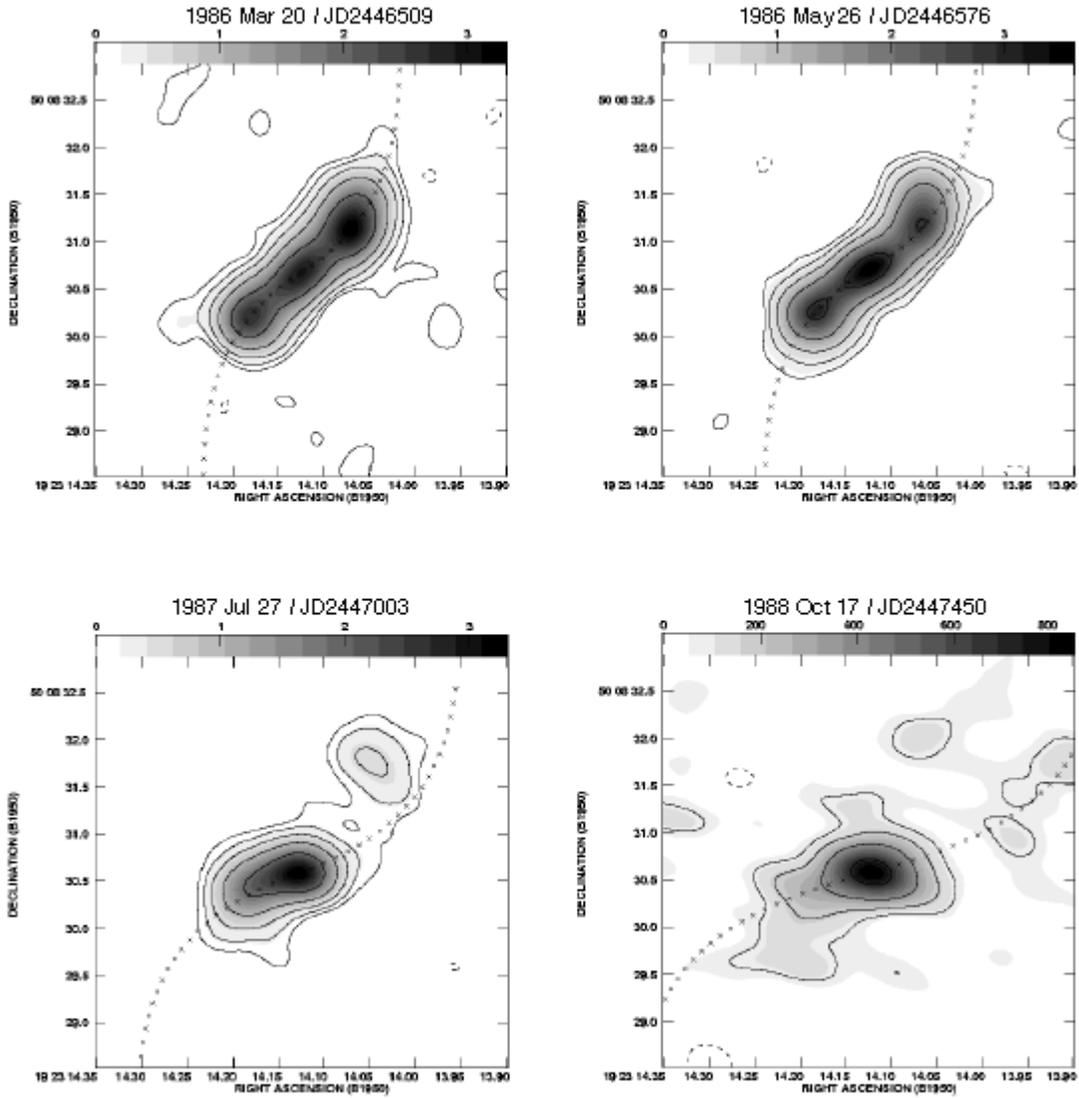}} 
\end{picture} 
\caption{Fitted ballistic material ejection model overlaid upon VLA
5GHZ radio maps (1986-1988).}
\label{fig:ballistic1}
\end{figure*}

\begin{figure*}
\begin{picture}(400,600)
\put(0,0) 
{\includegraphics{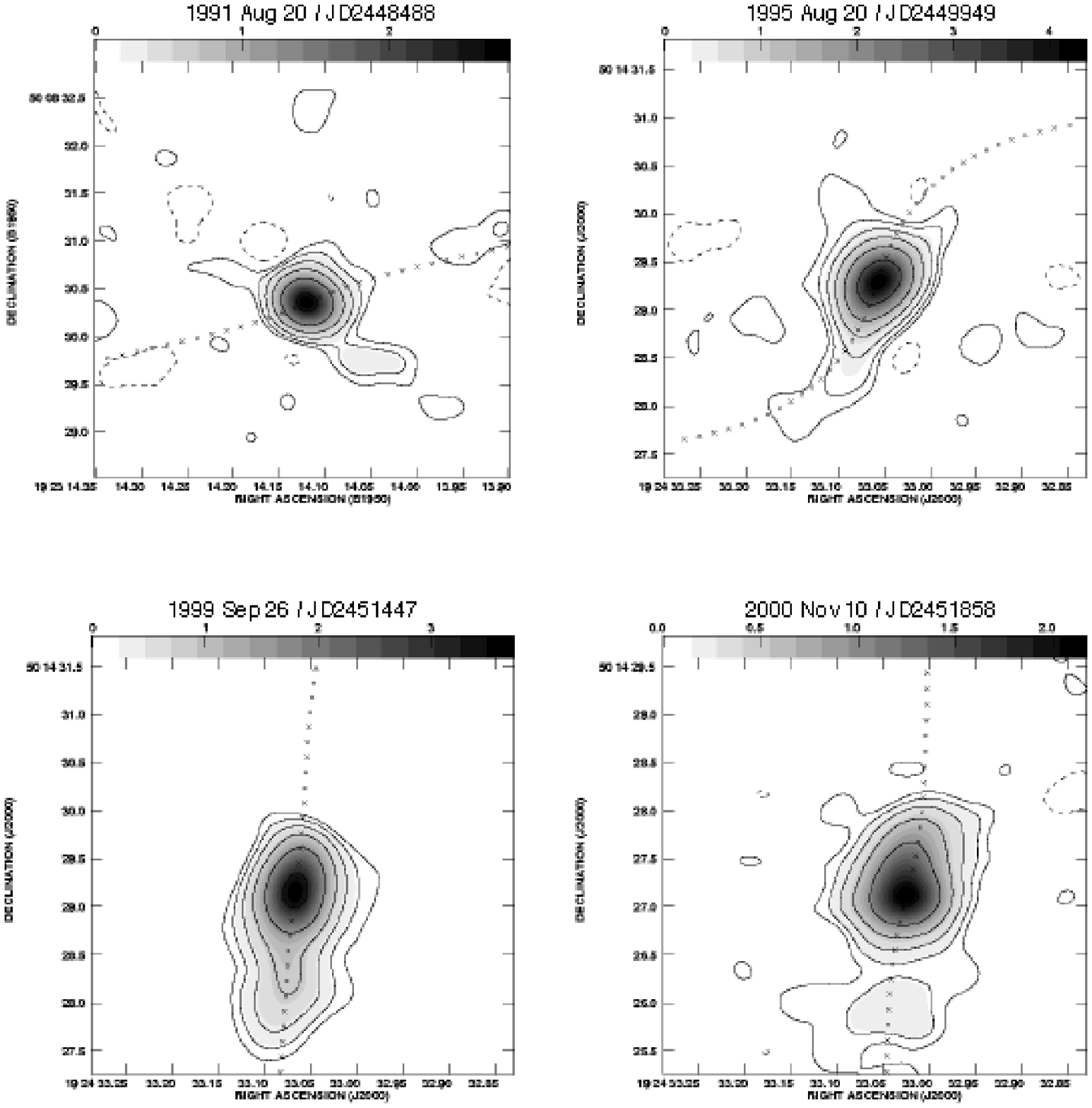}} 
\end{picture} 
\caption{Fitted ballistic material ejection model overlaid upon VLA
5GHZ radio maps (1991-2000).}
\label{fig:ballistic2}
\end{figure*}

The size of the initial simplex was set up so that it had parameters
centred around $P=5294$~days \citep{KHH93}, $T_0=$JD2446275
\citep{MM90a}, $\theta_{\rm mid}=140^\circ$, $\theta_0=30^\circ$
(Section \ref{sec:precinner}), $V=1230$~kms$^{-1}$ (Paper I) and
$i=90^\circ$. The process was allowed to run until the standard
errors of the evaluated fits became smaller than 0.01. The ``best--fit''
parameters from this run were then fed in as the starting values for a
second run. This process was repeated so that four runs were carried
out. The final set of parameters is shown in Table \ref{tab:sbm}. 
Results very similarb to this were found if the starting period 
was set to 8373 days, as found in  Section \ref{sec:precinner}. 
The resulting ephemerides, resulting from material being ejected in 
discrete bullets every 50 days  and following the model described above, 
are shown in Figures \ref{fig:ballistic1} and \ref{fig:ballistic2}.

\begin{table} 
\caption{Fitted parameters for the simple ballistic model of jet
ejection and precession.}
\begin{tabular}{|l |l |} 
\hline 
Parameter          &Value\\
\hline
$P$                &$6519 \pm 153$ days\\ 
$T_0$              &JD$2444038 \pm 128$ \\ 
$\theta_{\rm mid}$ &$140^\circ \pm 1^\circ$ \\
$\theta_0$         &$35^\circ \pm 1^\circ$ \\ 
$v_0$              &$1263 \pm 18$ km~s$^{-1}$\\ 
$i$                &$88^\circ \pm 1^\circ$ \\
\hline 
\end{tabular}  
\label{tab:sbm} 
\end{table}

The model fits reasonably well to the geometry seen in the maps,
especially in those resulting from observations closely following the
emergence of  the radio jet in 1986. In these images (JD2446509 and
JD2446576) the  jets are two--sided and well extended, making fitting
more robust.

In later epochs, when the jet is more one--sided, the curves still
fit the morphology seen in the radio maps well. The fact that such a fit is
achieved so long after the initial outburst (and following at least
two other outbursts) demonstrates that the behaviour of the bipolar
jets is not far from a simple ballistic model of the kind applied
successfully to SS433 and other sources.

\subsection{Disrupted jet model}

The greatest deviations of the simple model from the observations are
seen in the 1991 map (which has a low peak brightness and high
background noise level) and in the motion of the NW component in the
1987 and 1988 maps. If the component is the
same object in all of the maps, then it is moving in a fashion
entirely different to a simple, ballistic, bullet of material. By
inspection, it appears to be decelerating, perhaps due to interaction
with dense material.

The synchrotron theory put forward in Paper I required ejecta to be
impacting upon previously ejected stellar wind material, in order
to cause shocks and local condensations in the magnetic field. This 
might be expected to cause some degree of deceleration and disruption.

Direct evidence of deceleration is provided by optical spectroscopy
\citep{AS01}, which detects the velocity of material
within a few hundred R$_\odot$ of the central stars. Terminal
velocities of up to 2500~km~s$^{-1}$ are seen. Radio maps,
which observe the motion of material much further out (from a few AU
up to 100s of AU), never detect on--the--sky velocities of more than
1500~km~s$^{-1}$, assuming a distance to CH Cygni of 268~pc
\citep{RV98}. This is evidence of a large degree of deceleration
between these two scales.

The actual offsets of the apparently--slowed component from the
central radio peak at the different observing epochs are shown in
Table \ref{tab:nwoffset}.

\begin{table} 
\caption{Offsets of the anomalous northwestern component of the radio
jet of CH Cygni.}
\begin{tabular}{|l |l |l |l |l |l |} 
\hline  
Date    &Band &\multicolumn{2}{|l||}{Offset (mas)}&$\theta_{\rm PA}$ ($^\circ$) &$R$ (mas)\\
        &     &RA                 &Dec            &                             &\\ 
\hline 
2446509 &U    &$671 \pm15$        &$904 \pm 13$   &$323.4 \pm 1.0$              &$1130 \pm 10$\\ 
2446576 &U    &$546 \pm 47$       &$902 \pm 20$   &$328.8 \pm 2.8$              &$1050 \pm 30$\\
2447003 &C    &$773 \pm 75$       &$1195 \pm 29$  &$327.1 \pm 1.8$              &$1420 \pm 30$\\ 
2447450 &C    &$532 \pm 75$       &$1430 \pm 49$  &$339.6 \pm 3.2$              &$1530 \pm 50$\\ 
\hline
\end{tabular}  
\label{tab:nwoffset} 
\end{table}

A component moving ballistically, as described in Section
\ref{sec:ballistic}, would exhibit a constant position angle
($\theta_{\rm PA}$) and a radial distance from the core ($R$) that
increased  linearly with time. If there is simple deceleration,
$\theta_{\rm PA}$ would remain constant while $R$ changed according to
an environment--dependent model.

Compared with the $\sim 1230$~km~s$^{-1}$ velocity seen in the main
portion of the jets, the north western emission is almost stationary 
on the sky. It is
possible that the material has indeed almost stopped moving, perhaps
due to interactions with the surrounding environment. The jet may have
become entirely disrupted through shock--interactions and so would
no longer obey the simple ballistic model.

For a jet that has a time--variable ejection direction, such as a
precessing bipolar jet, shocks will form between the jet material and
the surrounding medium not only in the main direction of jet
propagation but also as a result of the transverse interactions
between the two regions \citep{JED87}. 

The transverse velocity of the jet becomes an important factor at large
distances from the point of ejection.

A colimated outflow will only survive until the transverse shock has crossed
the diameter of the jet. To simplify things, the approximation of \citet{ACR93a}
is used so that

\begin{equation}
t_{\rm d} \approx \sqrt{\frac{r_{\rm j}}{\beta v_0\Omega {\rm sin}\theta_0}}.
\label{eq:disrupttime}
\end{equation}

Most of these values can be estimated from work already carried out.
The jet radius, $r_{\rm j}$, can be estimated by measuring
the size of the anomalous region in the 1986 VLA U band map.
The region was found to be approximately
circular with a mean diameter of 0.18~arcsec. Assuming a distance of 
268~pc this corresponds to a linear size of $6.9\times 10^{12}$~m. The
initial jet ejection velocity was found in Paper I to be $1.23\times
10^6$~m~s$^{-1}$ which is in agreement with that found by fitting
to the the purely ballistic ejection model. The rate of precession
can be found from the fitted period of 6519 days to be $\Omega = 1.1\times
10^{-8}$rad~s$^{-1}$. The opening angle, $\theta_0$ was estimated
previously to be $35^\circ\pm 1^\circ$.

The main unknown is the relative density parameter,
$\beta$. Following \cite{ACR93a}, $\beta$ is assumed to
be constant, so that the ambient medium is entirely homogeneous.
This may not be the case, as a constant spherical outflow from 
a star is expected to produce an ambient density that varies as
$\rho_{\rm e} \propto r^{-2}$, and the presence of a jet would
be expected to cause disruptions in the local region. However, without
more knowledge of this region, a homogeneous medium is assumed for
simplicity. \cite{ACR93c} derived a ratio of $\rho_{\rm e}/\rho_{\rm j}
\approx 0.05$ from line ratios in the Herbig--Haro object HH 111
V. If this applies to CH Cygni, then $\beta$ is approximately
equal to 0.2.

Applying these parameters to Equation \ref{eq:disrupttime} gives a
value for the time taken for the disruptive shock to cross the jet as
$t_{\rm d} \approx 7\times 10^7$~s. The distance the material has
covered in this time (along the $x$ axis in Figure
\ref{fig:simplejet}) will be, for the parameters discussed above,
$x_{\rm d} = 311$ AU. The actual distance of the observed
anomalies in the radio jet in 1986--1988 is indeed around 300~AU.

To determine whether this disruption is occurring, direct
imaging of the shocked region is necessary. This has been done in the
case of the Herbig--Haro objects HH46/47 \citep{BR91} and HH34
\citep{BR92}. H$\alpha$ maps were subtracted from [S II] maps (at
wavelengths 6717\r{A} and  6731\r{A}) to produce images in which the
leading bow shock and trailing jet shock could be seen separately. The
ratio of the intensity of emission in the H$\alpha$ and [S II] was
then used to derive the relative densities in the jet material and
surrounding medium \citep{ACR93c}. Unfortunately, whilst H$\alpha$
emission maps of CH Cygni exist (Paper II), no maps in [S II] 
are available.

\subsection{Models with reflective symmetry}

If the geometry of the jets is a result of them being gravitationally
attracted to the outer giant, they would be expected to exhibit a
morphology in which the position angle of the material is a function
of distance from the ejection source, and in which there is reflective
symmetry about the central core \citep{LZ88}. Such a model was
applied successfully to 3C449 by \citet{RHL82}. Such a pattern does
not fit the observed extensions in  the CH Cygni system, implying that
the precession is due to effects within the collimating mechanism,
rather than post--ejection gravitational influences.

\subsection{Ballistic model with time--dependent ejection velocity}

CH Cygni is observed to undergo irregular outbursts with associated 
variable mass outflow rates. Such a state of affairs renders  invalid 
the assumption that what is being observed is a simple, continuous 
jet with constant ejection velocity. A jet that results from 
truly irregular outflows is difficult to model analytically using 
the methods described within this work.

Recent work, as described in Paper II, has discovered what may be two
types of mass outflow from the system. One is optically thin, at a
velocity measured spectroscopically to be 1200~km~s$^{-1}$ and the
other is optically thick and highly irregular. There is evidence that
the increase in mass--transfer from the cool component to the
compact component that occurs near to times of periastron leads to
higher luminosity of the compact component and, as a consequence,
greater radiatively driven mass outflow. 

Assuming a distance of 268~pc \citep{RV98}, the movement
between 1986 and 1988 shown in Table \ref{tab:nwoffset} represents
an on--the--sky velocity of only $\sim 200$~km~s$^{-1}$. Extrapolating
this motion backwards gives the time of ejection of this material,
assuming ballistic motion, as around December 1978. This time is
near the beginning of the large outburst that lasted between 1977
and 1984. Minor outflows have been observed spectrally during the 
outburst (\citealt{MH88}; \citealt{CMA80}) at velocities of
a few hundred km~s$^{-1}$. These may coincide with the ejection of
this anomalous packet of material.

Recent HST observations (from October 1999) made by \citet{RLC01} show 
the existence of loops of material in the [OII] nebula at a distance 
of $\sim 1.5$ and $\sim 5$ arcsec. Echelle spectroscopy of these loops 
places their kinematical axis at around 100~km~s$^{-1}$. 
Allowing for projection effects (i.e. assuming they are tilted at 
37$^\circ$ to the plane of the sky) their kinematical ages are 
approximately 15.2 years for the inner loop and 51 years for the outer 
one, giving the times of ejection of the loops as August 1984 and 
February 1949 respectively. The 1984 event fits in well with the
emergence of the radio jets first observed by \citet{ART86} and the
model proposed in this section, but no records of a 1949 outburst could
be found \citep{MM90a}.

\section{Possible causes of precession}

Precession of bipolar jets is a common phenomenon in astrophysics 
\citep{JF86}. The precession is usually attributed to motion of the 
collimating mechanism of the jets.

A simple collimating structure is a circumstellar torus or disk, such
as that postulated by \citet{JF83}. Flickering in the optical
brightness of CH Cygni has been observed on several occasions
(\citealt{GW68}; \citealt{MM90b}) and has been explained by the
orbital motion of a bright point on a disk surrounding the compact
component.

There are several mechanisms by which such a disk could be induced
to precess, and four of these scenarios are examined here. Only
one of these is able to cause precession at the rate
that is observed, given suitable estimates for the appropriate
parameters.

\subsection{Precession of a tilted accretion disk}

The gravitational influence of the nearby cool component, or perhaps 
the possible third star in the system, could force precession in this 
disk, and hence the jets. Following the work of \citet{DM80}, the angular 
speed of a torus or disk, with radius $r$, undergoing forced precession 
as a result of the influence of a companion is

$$
\Omega = -\frac{3}{4}\sqrt{\frac{G}{m_{\rm h}}}\frac{m_{\rm h}m_{\rm
c}}{m_{\rm h}+m_{\rm c}}\frac{r^\frac{3}{2}}{a^3}{\rm cos}\alpha
$$

\noindent where $m_{\rm h}$ is the mass of the hot component, $m_{\rm
c}$ is the mass of the cool component, $a$ is their separation and
$\alpha$ is the angle between the equatorial plane of the torus and
the orbital plane of the stars. Estimates for the parameters, from the
work of \citet{AS97}, are $m_{\rm h}=0.2$M$_\odot$,  $m_{\rm
c}=0.93$M$_\odot$ and $a=1.5$~AU. This relation assumes that
the disk is centred on the hot component, and the companion star
causing the forced precession is the cool giant on the 756 day orbit.
It would also be sensible to assume that $\alpha < 45^\circ$
\citep{MM90b} and $0 < r < a$. To obtain a precession period of
8373~days requires  a disk radius of around 0.7~AU. This is a very
large accretion disk, and it would be hard to imagine how such a disk
could form purely via wind accretion mechanisms, making this scenario
unlikely.

If instead it is assumed  that CH Cygni is a binary system on a 
5294~day orbit, with $a = 7.9$AU the disk radius is found to
be around 21~AU, around twice the size of the proposed semimajor 
axis of the outer giant's orbit.

\subsection{Slaved disk}

An attempt to explain the observed 164~day period precession of the
bipolar jets in SS433 was made by \citet{EPH80}, by assuming that
the compact component was surrounded by an accretion disk that
precessed at a rate governed by the apsidal motion of the parent 
star.

Using Equations 12, 13 and 16 of that paper, taking reasonable
stellar and orbital parameters of CH Cygni from \citet{AS97} and 
assuming that the coefficient of apsidal motion is 0.005 \citep{MS58}
suggests that the precession period of a slaved disk would be around 
12000 years. This result is a strong function of the radius of the cool 
component, so an increase in this star's size by a factor of 4 would 
reduce this precession period to about 15 years. 

To evaluate the possibility of such a model, precise eclipse
timings need to be made in the future.

\subsection{Irradiated disk}

An accretion disk that is irradiated by a central source can
become warped as a result of a torque that arises due to
radiation that is re--emitted from the disk \citep{ML97}.

Such a warped disk might be expected to precess. In contrast 
to the gravitationally induced precession mentioned above, this is 
``self precession'', in that no companion star is required. 
Following the work of \citet{ML96}, the period of this precession 
is of the same order as the time it takes for the warping to
develop in the disk. Using sensible parameters for the mass
of a possible accretion disk (around $10^{-7}$~M$_\odot$), this 
timescale is of the order of $\sim$2000 years, making this an 
unsuitable model for precession.

\subsection{Magnetic precession}

The inner region of an accretion disk surrounding a magnetic
star, such as a white dwarf \citep{MM90b}, is subject to
a twisting torque and a warping torque. These arise as a result 
of the interaction between the surface current of the disk and the 
horizontal magnetic field that is a feature of the dipole field of 
the star \citep{DL99}.

The warping torque arises through the twisting of the field that
threads the disk vertically. The precessing torque is a result of
the azimuthal screening current that occurs if the disk is
diamagnetic. In this case, the vertical field lines cannot
penetrate the disk. An azimuthal current is induced in the disk.
The radial component of the central star's magnetic field interacts
with this current to set up a vertical force acting on the disk.
This force is uneven across the disk, so a net torque results. This
torque acts to cause the disk (and hence the collimation mechanism of
the jets) to precess about the magnetic axis of the star.

\citet{DL99} gives the angular precession rate of 
such a disk in Equation 2.35 of that paper. Presently, most of the
required quantities are unknown, so only an order of magnitude
estimate is possible. The magnetic moment of the white dwarf was
estimated to be $2\times 10^{28}$~T~m$^3$ using the method of 
\citet{KAP85}, and the surface density of the disk was assumed to
be of the order 10~kg~m$^{-2}$ \citep{DNL88}. The angle between the 
disk axis and stellar rotation axis was taken to be equal to the 
opening angle of the precession cone, 35$^\circ$, found previously. 
The magnetic axis of the system was assumed to lie along the disk 
axis.

The magnetically induced precession period is then directly related
to the radial distance, $r$, of the warped torus from the white dwarf.
Since this varies as $r^7$, small changes in $r$ greatly affect the
period. A warped torus at around 2~R$_\odot$ from the star will 
precess with a period of around 6000 days.

A magnetic collimation and ejection mechanism for the jets gives  a
possible explanation for their one--sidedness \citep{GDC96}.  Jets
powered by a central magnetic rotator \citep{MM90b} can become inherently
asymmetric, if the solution to the magnetohydrodynamic equations for
each hemisphere of the star are asymmetric. Such conditions arise when
there is no change in the relative sign of the poloidal and toroidal
fields across the equator of the star. The different magnetic states
above and below the equator can then easily lead to a condition
in which only one supports the collimation and ejection of material 
via a jet \citep{JCW92}.

\section{Conclusion}

It is certain that the bipolar jets of CH Cygni exhibit precession,
with a period controlled by the motion of the mechanism resposible for
the collimation  of the outflow. In addition, the variability of the
mass--loss rate from the cool component causes the material to be
ejected at a rate that is highly variable (between $\sim 500$ and
$2000$km~s$^{-1}$), leading  to complications in predicting the
outflow geometry. The activity of the bipolar ejection seems to be
tied to the state of the hot--component,  with fast jets seen at or
soon after times of optical outburst. This activity then decreases
during times of quiescence.

Although the precession period found here is similar to the orbital
period of the outer giant in the system, the current time resolution
of the observations does not rule out shorter periods. A precession
that resulted from the motion of the inner, symbiotic, pair and had
a similar period to its 756~day orbit would need observations to
be taken at least every year for it to be detected.

The simple ballistic model fits the geometry of the nebula extremely
well. Knots seen in the larger--scale radio maps (the VLA in C, X and U
band) that do not fit the simple ballistic model can be explained as
regions of jet disruption caused by sideways motion of the ejected
material brought about as a result of the precession. The predicted
distance at which this disruption would occur is in close agreement
with the observations, although it relies upon several assumptions
about the nature of the ambient material. Confirmation of this model 
would require simultaneous optical images in both H$\alpha$ and [S II].

An alternative explanation is that the anomalous material was ejected
at a much slower velocity and at an earlier time than the majority 
of the gas in the bipolar nebula. This would indicate the presence
of a minor ejection event prior to the main ejection of the jets
first seen by \citet{ART86}. The variable ejection velocity model
discussed here and the Echelle spectroscopy of \citet{RLC01} suggest
the emission of material with a velocity of no more than a few 100s 
of km$^{-1}$ just before the main ejection of the high--speed
material. 

The cause of the precession is unknown but, given realistic estimates
for the masses and separation of the stellar components, warping of the 
collimating accretion disk by a magnetic white dwarf is the only mechanism
that can give precession periods similar to those found by
model fitting. The other possible causes give periods that are several
orders of magnitude too large. The magnetic explanation also allows
for the one--sided nature of the emission seen in all VLA radio
maps following the initial outburst in 1984.

\section*{Acknowledgments}

MC is supported by a grant from the Particle Physics and Astronomy
Research Council (PPARC).  The contribution of AS was supported by a
grant of the Slovak Academy of Science, number 2/1157/01.  The VLA is
operated by the National Science Foundation operated under cooperative
agreement with Associated Universities, Inc. MERLIN is a national
facility operated by the University of Manchester on behalf of PPARC.
Ultraviolet spectral data are based on INES data from the IUE satellite,
operated by ESA.

\label{lastpage}
\end{document}